\documentclass[12pt]{iopart}
\usepackage{graphicx}

\begin{document}

\title[Exact solution for the 2-dimensional Anderson localization]{Exact analytic
solution for the generalized Lyapunov exponent of the
2-dimensional Anderson localization}

\author{V N Kuzovkov\dag\ddag, W~von Niessen\ddag,
V Kashcheyevs\dag \ and O~Hein \ddag}

\address{\dag \ Institute of Solid State Physics, University of
Latvia, 8 Kengaraga Street, LV -- 1063 RIGA, Latvia}
\address{\ddag\ Institut f\"ur Physikalische und Theoretische Chemie,
Technische Universit\"at Braunschweig, Hans-Sommer-Stra{\ss}e 10,
38106 Braunschweig, Germany}

\ead{kuzovkov@latnet.lv}

\date{Received \today}
\begin{abstract}
The Anderson localization problem in one and two dimensions is
solved analytically via the calculation of the generalized
Lyapunov exponents. This is achieved by making use of signal
theory. The phase diagram can be analyzed in this way. In the one
dimensional case all states are localized for arbitrarily small
disorder in agreement with existing theories. In the two
dimensional case for larger energies and large disorder all states
are localized but for certain energies and small disorder extended
and localized states coexist. The phase of delocalized states is
marginally stable. We demonstrate that the metal-insulator
transition should be interpreted as a first-order phase
transition. Consequences for perturbation approaches, the problem
of self-averaging quantities and numerical scaling are discussed.
\end{abstract}

\submitto{\JPCM} \pacs{72.15.Rn, 71.30.+h}

\maketitle

\section{Introduction}

Frequently problems arise in science which involve both additive
and multiplicative noise. The first type is relatively easy to
handle with the help of the central limit theorem. The situation
changes dramatically with the appearance of multiplicative noise.
Famous examples are the Anderson localization, turbulence, and the
kicked quantum rotator among others. In this field results of an importance
comparable to the central limit theorem are still lacking.
Moreover, the approaches are in general numerical ones and
analytical tools are the rare exception.

We present such an analytic approach which permits to deal in
great generality with processes involving multiplicative and
additive noise even in the limit of strong disorder. In this paper
we apply the formalism to the famous Anderson localization in a
two-dimensional (2-D) disordered system which is one of the
paradigms of solid state theory.

The quantum mechanical consequences of disorder in solids
have first been revealed
by Anderson \cite{Anderson}. The Anderson model provides a
standard framework for discussing
the electronic properties of disordered systems, see reviews
\cite{Lee,Kramer,Janssen98}.
The nature of electronic states in the Anderson model depends strongly on
the spatial dimension $D$. It has been shown rigorously that
in one dimension (1-D) all states are localized at any level
of disorder \cite{Kramer,MottTwose}. The shape of these localized
wave functions is characterized by an asymptotic exponential decay
described by the Lyapunov exponent $\gamma$.
The most important results for  dimensions higher than one follow from the
famous scaling theory of localization \cite{Abrahams,Anderson2},
which assumes a single scaling parameter for
the dimensionless conductance $g$ or, equivalently,
the localization length $\xi= 1/ \gamma$.
The conclusion of the scaling theory
is that for $D \le 2$ all states are localized
at any level of disorder, while a delocalization (metal-insulator) transition
occurs for $D>2$ if the disorder is sufficiently strong.
A detailed review of the scaling theory for disordered
systems  can be found in \cite{Lee,Janssen98}.

The 2-D case still presents a problem, since there is no exact
analytical solution to the Anderson problem, and all numerical results
published so far rely on finite-size scaling \cite{Kramer,Zkarekeshev96}.
Recent studies \cite{Kantelhardt02}
have questioned the validity of the single parameter scaling theory,
including the existence of a finite asymptotic localization length
for $D=2$.
Additional boost of interest in the $2D$ Anderson model has been triggered
by experimental observations of Kravchenko et al. \cite{Kravchenko,Abrahams2}
of a metal-insulator transition in thin
semiconductor films, which contradicts the conventional
scaling theory. Moreover, recent experiments of Ilani et al.
\cite{Ilani1,Ilani2} can be interpreted in terms of the coexistence of
localized and delocalized states. These experiments are still being discussed
controversially.
The experimental reality is certainly more complex than the simple
tight-binding schemes used in the theoretical treatment so far and in particular
the electronic iteractions could play a role
in the above mentioned experimental situations. But nevertheless these results add doubts
to the status of the localization theory in 2-D. Before embarking on computational
schemes beyond the tight-binding approach, which necessarily lead to more
restricted system sizes and other approximations,
it appears advisable to try to solve as rigourously as possible the problem
in the tight-binding scheme. In the present controversial situation the first
step in resolving the conflict is thus in our opinion to consider
exact results that do not rely on the
scaling theory or small parameter expansions.

The starting point for the method presented in this paper is found
in the work of Molinari \cite{Molinari}, in which the Anderson
problem for the 1-D system is dealt with as a statistical
stability problem for the solutions $\psi_n$ of the tight binding
Hamiltonian in a semi-infinite system, $n\geq 0$. It was shown in
ref. \cite{Molinari} that the equations for the statistical moments of
the type $ \langle \psi^{2j}_n \rangle$ can be obtained
analytically (explicit solutions are given for $j=1,2$), which
enabled the author to derive exact generalized Lyapunov exponents.
We will show in the following that this approach can be further
generalized for systems of higher spatial dimensions. But it turns out to be
unavoidable to change again the mathematical tools for the treatment.
In the present investigation we use both for the 1-D and the 2-D case the tool
of signal theory abundantly used in electrical engineering, see e.g.\cite{Weiss}.
The basic idea in applying signal theory to the problem of Anderson localization
is to interpret certain moments of the
wave function as signals. There is then in signal theory a qualitative
difference between localized and extended states: The first ones
correspond to unbounded signals and the latter ones to bounded signals.
In the case of a metal-insulator transition extended states
(bounded signals) transform into localized states (unbounded signals).
Signal theory shows that it is possible in this case to find a
function (the system function or filter), which is responsible for
this transformation. The advantage of working with filters instead of the
signals themselves lies in the fact that the filters do not depend on initial
conditions in contrast to the signals. The existence of this transformation in
a certain region of disorder and energy simply means that
the filter looses its stability in this region. The meaning of an
unstable filter is defined by a specific pole diagram in the complex
plane. These poles also define a quantitative measure of
localization. Thus it is possible here to determine the socalled
generalized Lyapunov exponents as a function of disorder and energy.

The outline of the present article is as follows. In chapter 2 we treat the 1-D case
in detail describing also essential elements of signal theory. The theory for the
2-D problem is presented in chapter 3 and the results are given in chapter 4.
In the latter chapter also
the implications of the present approach for perturbation theory, the order
of the phase transition and the problem of self-averaging is discussed as well as numerical scaling.

\section{1-D case}
\subsection{Recursion relation}
We start for pedagogical and methodical reasons with the treatment
of the 1-D case. The aim of this section is to apply mathematical
tools which are new in the field of Anderson localization but well-known
from other fields and which may and do in fact prove useful
also for the higher dimensional cases.

We start from the Hamiltonian of the standard 1-D Anderson model
in the tight-binding representation
\begin{equation} \label{introduction_e1}
\mathcal{H}=\sum_n\varepsilon _n| n \rangle \langle n| + t \sum_n
\left[ |n \rangle \langle  n+1| + |n \rangle \langle n-1| \right
],
\end{equation}
where $ t $ is the hopping matrix element (units with t=1 are used
below) and the $ \varepsilon_n $ are the on-site potentials which
are random variables to simulate the disorder. The $ \varepsilon_n
$ are independently and identically distributed with existing
first two moments, $\left\langle \varepsilon _n\right\rangle =0$
and $\left\langle \varepsilon _n^2\right\rangle =\sigma ^2$. This
model is well investigated \cite{Kramer,MottTwose,Ishii} and it is
known that for any degree of disorder the eigenfunctions become
localized. This in turn means that $ \vert\psi_n\vert^{\delta} $
with $\delta>0$ is growing exponentially in the mean on a
semi-infinite chain where the rate of growth is described by the
Lyapunov exponent.

In the one-dimensional case one can compute the Lyapunov exponent
by different methods and under different assumptions \cite{Kappus}
or numerically  by means of the transfer matrix method which
results in solving a recurrence equation \cite{Pendry88}. The
disadvantage of these analytical methods is that they cannot be
easily extended to the 2-D and 3-D cases. In this paper we want to
show that one can construct an approach which is able to handle
all cases from a universal point of view.

It is well-known that the two-point boundary problem for a finite
difference analog to the stationary Schr\"{o}dinger equation with
homogeneous boundary conditions can be reduced to a Cauchy problem
with the one-side (initial) condition \cite{Thouless}
\begin{equation}\label{rand}
\psi_0 =0, \psi_1= \alpha.
\end{equation}
Due to the linearity of the equation the non-zero value of
$\alpha$ serves only as a normalization parameter. The asymptotic
behaviour of the solution is completely determined by the energy
$E$ and the level of disorder $\sigma$. The presence of a boundary
permits one to rewrite the Schr\"{o}dinger equation
\begin{equation} \label{schro}
\psi _{n+1}+\psi_{n-1}=(E-\varepsilon _{n}) \psi _{n}
\end{equation}
in the form of a recursion relation for the grid amplitude of the
wave function $ \{ \psi_n \} $ ($n=1,2, \ldots$):
\begin{equation} \label{recursion}
\psi_{n+1} = (E-\varepsilon_n)\psi_n - \psi_{n-1}.
\end{equation}
Statistical correlations in this equation are separated: it is easy
to see that in a formal solution of the recursion relation the
amplitude $\psi_{n+1}$ depends only on the random variables
$\varepsilon_{n^{\prime }}$ with $n^{\prime }\leq n$, which is the
causality principle. This observation has been
extensively used by Molinari\cite{Molinari} in the solution for the 1-D case. We
emphasize that this property is fundamental as a key for obtaining
an exact solution to the Anderson problem for all dimensions. Note
that both amplitudes $\psi_n$ and $\psi_{n-1}$ on the r.h.s.\ of
eq.\ (\ref{recursion}) are statistically independent of
$\varepsilon_n$.

The equation for the first moment of the random variable $\psi_n$
does not include the parameter $\sigma $:
\begin{equation} \label{rec1}
\left\langle \psi _{n+1}\right\rangle =E\left
 \langle \psi
_{n}\right\rangle -\left\langle \psi _{n-1}\right\rangle .
\end{equation}
The ansatz  $\left\langle \psi_n \right\rangle=\lambda^n$ results
in the $\lambda_{1,2} $ which satisfy the equation $\lambda +
{\lambda}^{-1} = E$. Energies with $|E | < 2$, where
$|\lambda_{1,2}|=1$ and $\lim_{n \to \infty} \langle \psi _n
\rangle < \infty $, obviously coincide with the band of
delocalized states in a perfect chain ($\sigma=0$).

\subsection{Lyapunov exponents as order parameter}

A bounded asymptotical behaviour of the first moment at any level
of disorder indicates that there are always physical solutions
inside the band $|E | < 2$. Further information about the
character of these states (localized / delocalized) can be gained
by considering the other moments. For one-dimensional models with
random potential the eigenstates are always exponentially
localized \cite{MottTwose}. The natural quantities to investigate
is therefore the Lyapunov exponent $\gamma_0$
\begin{equation}\label{gamma_0}
\gamma_0=\lim_{n\rightarrow \infty }\frac 1{n}\left\langle \ln
|\psi _n| \right\rangle
\end{equation}
or its generalization \cite{Molinari,Pendry88}
\begin{equation}
\gamma_{\delta}=\lim_{n\rightarrow \infty }\frac 1{n\delta}\ln
\left\langle | \psi _n|^{\delta} \right\rangle .
\end{equation}
The generalized exponents have been studied extensively by Pendry et al.
(see ref. \cite{Pendry88} and references given there),
in a systematic approach based on the symmetric group. The method involves
the construction of a generalized transfer matrix by means of direct
products of the transfer matrices, followed by a reduction of the matrix size.
This generalized transfer matrix produces the average values of the required
power of the quantity under consideration.

The concept of Lyapunov exponent $\gamma$ and the corresponding
localization length $\xi$ describe a statistical stability of
solutions of the tight-binding equations. Anderson localization as
a metal-insulator transition is a typical critical phenomenon.  To
determine the phase-diagram of the system we need only make use of the
qualitative aspect of the Lyapunov exponent. Two phases differ
qualitatively: $\gamma \equiv 0$ for conducting states (metallic
phase) and $\gamma \neq 0$ for localized states (insulating
phase).

The Lyapunov exponent $\gamma$ is a typical order parameter
\cite{Stanley}. It is useful to check whether this kind of
relation is valid for other systems like ferromagnets and
ferroelectrics too. Here the order parameters are the
magnetization $M$ and the polarization $P$, respectively. At high
temperatures and zero external field these values are $M,P\equiv
0$ (paramagnetic or paraelectric phase). At lower temperatures,
however, spontaneous magnetization and polarization arise,
$M,P\neq 0$. The order parameter is zero for one phase and
becomes nonzero for the other phase. It is well-known that there is no
unambiguous definition of an order parameter \cite{Stanley}.
This holds also in our case, many different
definitions are possible (see e.g. ref. \cite{Molinari} and literature
cited there). Every exact Lyapunov
exponent either via the log-definition ($\gamma_0$) or via the
$\delta$-definition ($\gamma_{\delta}$) gives this property.
This permits us to consider a transition from the
quality $\gamma \equiv 0$ to the quality $\gamma \neq 0$ as a critical
point, where all moments $\left\langle | \psi _n|^{\delta}
\right\rangle$ diverge for $n \rightarrow \infty$ simultaneously.
If on the other hand we are interested in the values themselves of
the Lyapunov exponents $\gamma \neq 0 $, they always depend on the
definition. If one prefers a particular
definition, this can only be by agreement and always remains quite
arbitrary.

In the 1-D case we find the metallic phase ($\gamma=0$) only for
$\sigma=0$; even for infinitesimally small disorder one has $\gamma
\neq 0$ and only the insulating phase exists. The critical point is
therefore $\sigma=\sigma_0=0$ independently of the value of the energy $E$,
and the phase-diagram is trivial. This situation is
typical for a critical phenomenon in a 1-D system with
short range interaction \cite{Stanley}: all 1-D
systems (e.g. 1-D Ising-model) do not possess a true
phase transition at a finite value of the parameter
(temperature in the Ising model or disorder $\sigma$ in
the Anderson model); one of the phases exists only at a point.
One also says \cite{Stanley}, that 1-D
systems do not possess a phase transition.

The advantage of the new  $\delta$-definition rests in the possibility
to play with the parameter $\delta$ which gives a new
degree of freedom. In quantum theory there are many examples, where
a problem is insoluble in the Schr\"odinger picture, but looks rather simple in
the Heisenberg picture. The same holds for representations in coordinate
or momentum space. It is always useful to transform the mathematics
in such a way that the problem becomes soluble. And nothing else is done here.
For $\gamma_0$ (log-definition)
the problem remains analytically insoluble.  Molinari \cite{Molinari} was the first
to show that with a generalisation of the definition,
and for special values of $\delta=2,4,...$, a
simple analytical (algebraic) investigation becomes possible.

Molinari\cite{Molinari} has not only considered a standard
definition via "the log of the wave function" i.e. $\gamma_0$, but
also a set of so-called generalized Lyapunov exponents
$\gamma_{2j}$, $j=1,2,...$ ($\gamma_2$ for the square of the wave
function). In general all the $\gamma_{2j}$ are quite different
parameters. However, it is important that a transition from the
quality $\gamma \equiv 0$ to the quality $\gamma \neq 0$ is
simultaneous, all $\gamma_{2j} \equiv 0$ for $\sigma=0$ and
$\gamma_{2j} \neq 0$ for a disordered system. As has already been
established \cite{Kramer,Kappus}, in the limit of small disorder
different definitions of the Lyapunov exponent or localization
length give values differing only by an integer factor. This
property is again a signature for a critical phenomenon. In the
vicinity of the critical point $\sigma=\sigma_0=0$ as common for
all critical phenomena and only one scale dominates. Thus Molinari
has shown that for the 1-D Anderson problem all $\gamma_{2j}
\approx (1+j) \gamma_0$ for $\sigma \rightarrow 0$ (this is proven
for $j=1,2,3$), where $\gamma_0=\sigma^2/2(4-E^2)$ is taken from
other investigations\cite{Kramer,Kappus,Thouless}, the only
difference being the numerical cofactor.

This means that even in 1-D there are other quantities besides the
logarithm of the wave function which can be used for the analysis.

\subsection{Equations for second moments}

In order to obtain the phase diagram it is (based on the above
discussion) sufficient to choose a particular and convenient value of
the parameter $\delta$, e.g. $\delta=2$, the second moments.

It is shown in ref. \cite{Molinari} that the calculation of the higher
moments, $\langle \psi^{2j}_n \rangle$ with $j>1$, is important
for determining the shape of the distribution of $|\psi_n|$, but
at the same time the higher moments of the on-site potentials
beyond the second one must be considered. We will restrict
ourselves here to the pair moments only. Then the initial full
stochastic problem eq. (\ref{recursion}) can be mapped onto an
exactly solvable algebraic problem, in which the random potentials
are characterized by a single parameter $\sigma$.

Two points  should be mentioned: (i) We are at present not
interested in the shape of the distribution which is influenced by
the higher moments of the on-site potentials but in the problem of
localization (phase-diagram); (ii) In the analysis of the moments
of the amplitudes the localization of states finds its expression
in the simultaneous divergence of the even moments for $n
\rightarrow \infty$. Because the second moment depends only on the
parameter $\sigma$ this means that the critical properties are
completely determined by $\sigma$.

We are interested in the mean behavior of $ \psi^2_n $ which
follows from eq.\ (\ref{recursion}) as:
\begin{eqnarray} \label{secondmoment_e2}
\langle \psi_{n+1}^2 \rangle = \langle[(E-\epsilon_n) \psi_n -
\psi_{n-1}]^2 \rangle \nonumber \\ = (E^2 + \sigma^2) \langle
\psi_n^2 \rangle - 2 E \langle \psi_n \psi_{n-1} \rangle + \langle
\psi_{n-1}^2 \rangle.
\end{eqnarray}
In the derivation of eq.  (\ref{secondmoment_e2}) the mean of the
product of uncorrelated quantities was replaced by the product of
the means. The resulting equation is open ended, but the new type
of the means $ \langle \psi_n \psi_{n-1} \rangle $ can be easily
calculated:
\begin{eqnarray} \label{secondmoment_e3}
\langle \psi_n \psi_{n-1} \rangle &=& \langle [(E-\epsilon_{n-1})
\psi_{n-1} - \psi_{n-2}] \psi_{n-1} \rangle \nonumber \\&=& E
\langle \psi_{n-1}^2 \rangle - \langle \psi_{n-1} \psi_{n-2}
\rangle.
\end{eqnarray}
Let us rewrite these equations using $x_n= \langle\psi^2_n\rangle
$ and $y_n=\langle \psi_n \psi_{n-1} \rangle $:
\begin{eqnarray}
    x_{n+1} &  = & (E^2 + \sigma^2) \,
    x_n - 2 E y_n + x_{n-1}, \label{relations}
    \\
    y_n & = & E \, x_{n-1} - y_{n-1}. \label{relations2}
\end{eqnarray}
The initial conditions are:
\begin{equation} \label{inicond}
    x_0=0 , x_1= \alpha^2, y_0=0.
\end{equation}
Let us summarize the intermediate results up to this point. The
causality principle has led to a set of linear algebraic
equations for the second moments of the random field instead of an
infinite hierarchy of equations that couple the second moments
with the third ones etc. The set of equations is closed, but
it does not include all possible second moments of the type
$\langle \psi_n \psi_{n^{\prime }} \rangle$ (this fact is
irrelevant in the search for localization criteria). At this level
the on-site potentials are also characterized by a single second
moment $\sigma^2$ only, which implies that the shape of the distribution
does not matter for localization. Information on higher moments is
a crucial step for other models, e.g., in turbulence and
econophysics because the third and the fourth moments are linked
to skewness and kurtosis of the distribution which are interesting
features in these fields. Equations for the higher moments can be
easily constructed using the same causality
property\cite{Molinari}, but in the treatment of the Anderson
problem we restrict ourselves to the second moment only. The set
of equations is exact in the sense that no additional
approximations were made in its derivation. Therefore, we conclude
that at the given point the stochastic part of the solution to the
Anderson problem is completed and one has to deal further only
with a purely algebraic problem.

The set of equations (\ref{relations}), (\ref{relations2}) can be
solved by different methods of linear algebra.

\subsection{Transfer matrix}

We employ first a simple matrix technique. The eqs.\
(\ref{relations}), (\ref{relations2}) can be rewritten in the form
\begin{equation} \label{matform}
    w_{n+1} = T \, w_n \, ,
\end{equation}
where the vector $w_n$ and the transfer matrix $T$ are
respectively
\begin{equation}
     w_n  = \left (
    \begin{array}{c}
        x_{n+1} \\ x_{n} \\ y_{n+1}
         \end{array} \right ) \, , \,
    T = \left (
    \begin{array}{ccc}
        E^2 + \sigma^2 & 1 & -2 E \\
        1 & 0 & 0 \\
        E & 0 & -1
        \end{array} \right ) \, .
\end{equation}
Initial conditions transform into $w_0=\{ \alpha^2, 0, 0 \}$.

An explicit formula for $w_n$ is derived by diagonalizing the
transfer matrix $T$. The characteristic equation of the eigenvalue
problem for the $T$ matrix is
\begin{eqnarray} \label{chareq}
\mathcal{D}(\lambda)=0
\end{eqnarray}
with function $\mathcal{D}(z)=z^3 - (E^2 + \sigma^2 -1) z^2 + (E^2
- \sigma^2 -1) z  -1$.

The solution of the cubic characteristic equation (\ref{chareq}) can
be given explicitly in radicals (Cardan's formula). Corresponding
expressions are known and not presented here. It is significant
that one of the eigenvalues (let it be $\lambda_1$) is always real
and $\lambda_1\geq 1$. Other solutions, $\lambda_2$  and
$\lambda_3$, are either complex conjugate to each other or both
real and satisfy $| \lambda_{2,3} | \leq 1$. These properties
result from the fact that the coefficients $E^2 $ and $\sigma^2$
are non-negative.

The solution of eq.\ (\ref{matform}) reads as follows:
\begin{eqnarray}
      w_n & = & U \cdot
      \Lambda(n) \cdot
       U^{-1} w_0,
  \end{eqnarray}
where $\Lambda(n)$ is a diagonal matrix containing the n-th power
of $\lambda_i$ and $U$ is the eigenvector matrix. The resulting
exact formula for $x_n$ is
\begin{eqnarray} \label{x}
      x_n / \alpha^2 & =  & \frac{\lambda_1^n (1 + \lambda_1)}
      {(\lambda_2-\lambda_1)(\lambda_3-\lambda_1)} \nonumber + \\
      & + & \frac{\lambda_2^n (1 + \lambda_2)}
      {(\lambda_1-\lambda_2)(\lambda_3-\lambda_2)}  + \\
      & + & \frac{\lambda_3^n (1 + \lambda_3)}
      {(\lambda_1-\lambda_3)(\lambda_2-\lambda_3)} \nonumber\, .
   \end{eqnarray}

Thus we have the full algebraic solution for any energy $E$ and
any degree of disorder $ \sigma$. From the functional form of the
eq. (\ref{x}) one can see that the roots of eq.\ (\ref{chareq}),
$\lambda_i$, give us the Lyapunov exponents $\gamma$ of the
problem, $\lambda=\exp(2\gamma)$, so the judgement on a
localization transition can be done immediately after arriving at
eq.~(\ref{chareq}). Electronic states satisfying the inequality
$\max | \lambda_{i} | = \lambda_{1} > 1$ correspond to
localization.

\subsection{Z-transform}
An alternative solution makes use of the
so-called Z-transform. This is used mainly in Electrical Engineering for
discrete-time systems and we suggest as publicly available source
\cite{Weiss}. The Z-transform of the quantities $x_n$ and $y_n$ to
functions $X(z)$ , $Y(z)$ is defined by

\begin{eqnarray}\label{XY}
X(z)=\sum_{n=0}^{\infty}\frac{x_n}{z^n} ,
Y(z)=\sum_{n=0}^{\infty}\frac{y_n}{z^n} .
\end{eqnarray}
The inverse Z-transform is quite generally defined via countour
integrals in the complex plane
\begin{eqnarray}\label{count}
x_n=\frac{1}{2\pi i}\oint X(z)z^n \frac{dz}{z} .
\end{eqnarray}
We further on need the following properties: for the Z-transform
\begin{eqnarray}\label{shift}
x_{n+n_0}\Rightarrow z^{n_0}X(z) ,
\end{eqnarray}
and for the inverse Z-transform
\begin{eqnarray}\label{lambda}
\frac{z}{z-\lambda} \Rightarrow \lambda^n .
\end{eqnarray}

In this way eqs. (\ref{XY}) and (\ref{shift}) translate  eqs.
(\ref{relations}), (\ref{relations2}) into a system of two coupled
linear equations for the two unknowns $X$ and $Y$
\begin{eqnarray}
zX &=& (E^2+\sigma^2)X - 2 E Y +
z^{-1}X + \alpha^2 \label{euk2_ein1_e3} \\
Y &=& Ez^{-1}X -z^{-1}Y
\end{eqnarray}
This is easily solved for $X(z)$
\begin{equation} \label{Xz}
X(z)= \frac{\alpha^2 z(z+1)} {\mathcal{D}(z)}
\end{equation}
where the function $\mathcal{D}(z)$ is defined above. The inverse
Z-transform gives us again eq.(\ref{x}). It is easy to see that the
solution of the characteristic equation (\ref{chareq})
for the transfer matrix is equivalent to the determination of the poles
of the $X(z)$ function. The eigenvalues $\lambda_i$, i.e. the poles,
determine according to eq. (\ref{lambda}) the
asymptotic behaviour of the solution $x_n$ for $n \rightarrow
\infty$. (This is a simplified example for the general relation).

\subsection{Signal Theory}

Let us start with the following definitions. Let
\begin{equation} \label{X00}
X^{(0)}(z) = \alpha^2 \frac{ (z+1)} {(z-1)} \frac{z}{(z+1)^2-zE^2}
\end{equation}
describe an ideal system ($X^{(0)}(z)=X(z)$ for $\sigma \equiv 0$).
This function is independent of the parameter $\sigma$.
For $\sigma\neq 0$
\begin{equation} \label{XH}
X(z)=H(z)X^{(0)}(z) ,
\end{equation}
with
\begin{eqnarray} \label{H0}
H(z)=\frac{(z-1) [(z+1)^2 -zE^2]}{\mathcal{D}(z)}.
\end{eqnarray}
Note that the boundary conditions (parameter $\alpha$) influence
only the function $X^{(0)}(z)$. The function $H(z)=1$ for $\sigma=0$.

Eq. (\ref{XH}) possesses quite a remarkable structure which is
better interpreted in the context of signal theory \cite{Weiss},
which makes intensive use of the Z-transform.  Let us define the
system input as $X^{(0)}(z)$ (it characterizes the ideal
system), the system output as $X(z)$ (the disordered
system), then the function $H(z)$ is the system function
or filter. The inverse Z-transform gives \cite{Weiss}
\begin{equation}\label{x_n}
x_n=\sum_{l=0}^nx_l^{(0)}h_{n-l} .
\end{equation}

Signal theory is not very crucial for the $D=1$ case,
because the solution, eq.(\ref{Xz}), looks very simple and the inverse
Z-transform is always possible. For $D>1$, however,
the use of signal theory is exceedingly important
because the corresponding solution $X(z)$ can be a very complicated function
and an inverse transform may be impossible to find.
In the present case it is, however, completely sufficient
to investigate the analytic properties of the filter function $H(z)$, i.e.
the poles $z=\lambda_i$, because the poles determine uniquely the properties
of the system. The essential idea is very simple.
In the band $| E | < 2$ all wave amplitudes (and input
signals $x_n^{(0)}$) are bounded. Output signals $x_n$ are
unbounded only under the condition that the filter $h_n$ is
unbounded for $n \rightarrow \infty$. This property depends on the position
of the poles $\lambda_i$ in the function $H(z)$.

It is known \cite{Weiss} that the filter $H(z)$ can be
characterized by a pole-zero diagram which is a plot of the
locations of the poles $\lambda_i$ and zeros in the complex
z-plane. Since the signals $x_n^{(0)}$ and $x_n$ are real, $H(z)$
will have poles and zeros that are either on the real axis, or
come in conjugate pairs. For the inverse Z-transform
$H(z)\Rightarrow h_n$ one has to know the region of
convergence (ROC).  As follows from physical reasons we are
interested only in causal filters ($h_n=0$ for $n<0$)
that have always ROCs outside a circle that intersects the pole
with $max{|\lambda_i}|$.  A causal filter is stable
(bounded input yields a bounded output) if the unit
circle $|z|=1$ is in the ROC.

To give an example. We start the analysis of the solution with the case
of the ideal system, $\sigma=0$. All solutions in the band $| E |
< 2$ (defined as a region with asymptotically finite first moment)
are delocalized with $\lambda_1= 1$, $\lambda_{2,3}=e^{\pm i
2\varphi}$, where $\varphi = \arccos(|E| /2)$. The inverse Z-transform
gives us $x_n=\alpha^2 \sin ^2(\varphi n)/\sin ^2(\varphi )$.
In this case the filter $H(z)=1$ and the ROC of this filter is the
full commplex z-plane. The ROC includes the unit circle, the
filter $H(z)$ is thus stable, which means the
delocalization of all states.

For $\sigma \neq 0$ we always have $\lambda_1>1$. The ROC corresponds
to the region $|z|\geq \lambda_1$, and the unit circle $|z|=1$
lies outside the ROC. A filter $H(z)$ is unstable, in
other words, this is simply the localization of all states.

\section{2-D case}\label{2D}

\subsection{Recursion relation}
Consider a 2-D  lattice with one boundary. The layers of this
system are enumerated by an index $n=0,1, \ldots$ starting from
the boundary, and the position of an arbitrary lattice site in a
particular layer is characterized by an integer $m \in (-\infty ,
+\infty )$. The presence of a boundary permits one to rewrite the
Schr\"{o}dinger equation
\begin{equation} \label{schroedinger}
\psi
_{n+1,m}+\psi_{n-1,m}+\psi_{n,m+1}+\psi_{n,m-1}=(E-\varepsilon
_{n,m}) \psi _{n,m}
\end{equation}
in the form of a recursion relation for the grid amplitude of the
wave function $ \{ \psi_{n,m} \} $ ($n=2,3, \ldots$):
\begin{equation} \label{recursion D}
\psi_{n,m}=-\varepsilon_{n-1,m}\psi_{n-1,m}-\psi_{n-2,m}+\mathcal{L}\psi_{n-1,m}
.
\end{equation}
For the sake of a compact notation an operator $\mathcal{L} $ is
introduced which acts on the index  $m$ according to the equation
\begin{eqnarray}\label{L}
\mathcal{L}\psi_{n,m}=E\psi_{n,m} -\sum_{\mu=\pm 1}\psi_{n,m+\mu}
.
\end{eqnarray}
The lattice constant and the hopping matrix element are set equal to
unity. The on-site potentials $ \varepsilon_{n,m} $ are
independently and identically distributed with existing first two
moments, $\left\langle \varepsilon _{n,m}\right\rangle =0$ and
$\left\langle \varepsilon _{n,m}^2\right\rangle =\sigma ^2$.
Eq.(\ref{recursion D}) is solved with an initial condition
\begin{equation}\label{rand D}
\psi _{0,m}=0,  \psi _{1,m}=\alpha_{m}.
\end{equation}
It turns out to be convenient to consider the index $n$ not as a
spatial coordinate, but as discrete time. Then eq. (\ref{recursion
D}) describes the time evolution of a $D-1=1$ dimensional system.
It is easy to see that in a formal solution of this recursion
relation the amplitude $\psi_{n,m}$ depends only on the random
variables $\varepsilon_{n^{\prime },m^{\prime }}$ with $n^{\prime
}< n$ (causality). We encounter a very important feature
in eq.\ (\ref{recursion D}): grid amplitudes on the r.h.s.\ are
statistically independent with respect to $\varepsilon_{n,m}$.

\subsection{Implications of signal theory}\label{signal}

We are going to generalize further the result of ref. \cite{Molinari}
that the set of equations for a certain combination of pair
moments is self-contained and can be solved analytically.

The divergence of the moments, which is caused by the localization,
is the basis for the existence of Lyapunov exponents for the $n$-direction,
which is generally a functional $\gamma[\alpha_m]$ of the field
$\alpha_m$. The well-known idea to define the fundamental Lyapunov
exponent for the problem of Anderson localization as a minimal one
$\gamma=\min \{ \gamma[\alpha_m] \}$, is an algorithm but not a
general definition, because a fundamental quantity is independent of the
initial condition. The proper definition is possible in the framework
of signal theory \cite{Weiss}.

We define the solutions of the equations for the second moments
with disorder ($\sigma >0)$ and without it ($\sigma=0$) as $x$
(system output) and $x^{(0)}$ (system input). Because these
equations are linear, there exist an abstract linear operator
$\hat{h}$ (system function or filter), which transforms one
solution into the other one, $x=\hat{h}x^{(0)}$. It is important
that the initial conditions $\alpha_m$ only determine the signals,
the filter on the other hand is a function of the disorder
$\sigma$ only. Divergence of the moments (unbounded output for
bounded input) simply means that the filter is
unstable\cite{Weiss}.

This approach utilizing the concept of the system function is a
general and abstract description of the problem of localization.
Instead of analyzing the signals $x$ which have restricted physical
meaning in the present context because of the chosen normalization
we study the filter $\hat{h}$ with properties described by
generalized Lyapunov exponents. Then e.g. delocalized states
(bounded output) are obtained by transforming the physical
solutions inside the band $|E | < 4$ (bounded input) provided that
the filter $\hat{h}$ is stable.

The transformation $x=\hat{h}x^{(0)}$ is not only valid for
individual signals, but also for linear combinations of these
signals. Let us regard an ensemble of initial conditions in eq.
(\ref{rand D}) which is obtained by trivial translation in
$m$-space, $\alpha^{\prime}_m=\alpha_{m+m_0}$. Translation
generates \textit{physically} equivalent signals with identical
Lyapunov exponents $\gamma[\alpha_m]$. A linear combination of
these signals also has this same value $\gamma[\alpha_m]$. We
construct a linear combination from all such signals with equal
weights (this corresponds simply to an average $\langle ...
\rangle _{0}$ over all possible translations $m_0 \in (-\infty ,
+\infty )$).

We here start from the basic fact that the determination of the
phase-diagram (the fundamental topic of the paper) requires only
the Lyapunov exponents. Signals $x$  are only a means to arrive there.
Consequently it is possible to make certain operations with the
signals, but under the strict condition that these operations have
no influence on the Lyapunov exponents $\gamma[\alpha_m]$.

Next we define a full averaging over
random potentials and over the ensemble of translations in
$m$-space. This latter averaging correponds to the construction of
the linear combinations discussed above. Full averaging restores
the translational invariance along the $m-$axis, which appears
e.g. in $m-$independent diagonal elements, $(n,m) =
(n^{\prime},m^{\prime })$, in the set of the second moments of the
type $ \langle \psi_{n^{\prime },m^{\prime }} \psi_{n,m}  \rangle
$ and $\langle \psi^2_{n,m} \rangle = x_n$. We further on regard
$x_n$ as a one-dimensional signal. If we succeed to solve the
equations for $x_n$ then we also find not only the Lyapunov
exponent $\gamma[\alpha_m]$ but simultaneously the projection of
the abstract operator on the one-dimensional space,  $\hat{h}
\rightarrow h_n$,  because for one-dimensional signals the
convolution property eq.(\ref{x_n}) exists \cite{Weiss}. The filter
$h_n$ possesses the same fundamental information as the abstract
filter $\hat{h}$ and can easily be analyzed, because signal theory
provides a definite mathematical language for this aim (see
above). We emphasize here that we do not reduce the problem to
a one-dimensional one; it remains two-dimensional. This is
quite apparent from the equations below which contain two
spatial variables, $n$ and $s=m-m'$, the distance along the m-axis.

\subsection{Second moments}

After the full averaging as defined above the moment $\langle
\psi_{1,m^{\prime }} \psi_{1,m} \rangle = \langle
\alpha_{m^{\prime}} \alpha_m \rangle _{0} =\Gamma_s $ (where
$s={m-m^{\prime}}$) transforms the initial condition by replacing
the field $\alpha_{m}$ by its property correlation function
$\Gamma_s$ (it is assumed that the field itself and its
correlation function are finite). The initial condition for  $x_n$
reads as $x_0 = 0, x_1 = \Gamma_0 $.

Non-diagonal elements, $(n,m) \neq (n^{\prime},m^{\prime })$,
depend on the difference $s=m-m^{\prime } $. In the following we
will need only three types of moments $ f_{n,s}^{\nu}$  with
$n^{\prime}=n-\nu$, $\nu=0,1,2$. Let us denote them by
$a_{n,s}=f_{n,s}^{0}$, $b_{n,s}=f_{n,s}^{1}$, and
$c_{n,s}=f_{n,s}^{2}$:
\begin{eqnarray}\label{Ma}
a_{n,s}=<\psi_{n,m^{\prime}}\psi_{n,m}> ,\\\label{Mb}
b_{n,s}=\frac{1}{2} \left
\{<\psi_{n-1,m^{\prime}}\psi_{n,m}>+<\psi_{n-1,m}\psi_{n,m^{\prime}}>
\right \} ,\\\label{Mc} c_{n,s}=\frac{1}{2} \left
\{<\psi_{n-2,m^{\prime}}\psi_{n,m}>+<\psi_{n-2,m}\psi_{n,m^{\prime}}>
\right \}  .
\end{eqnarray}
The corresponding initial conditions are $a_{0,s} = 0, a_{1,s} =
\Gamma_s$; $b_{0,s} = b_{1,s} = 0 $; and $c_{0,s} = c_{1,s} = 0 $.
Since the definition of $a_{n,s}$ for $s=0$ coincides with $x_n$,
we have the boundary condition
\begin{equation}\label{boundary}
a_{n,0} = x_n .
\end{equation}
The moments $f_{n,s}^{\nu}$ can be calculated directly from the
definition, eqs.(\ref{Ma})-(\ref{Mc}). Let us consider e.g.
$b_{n,s}$. In order to calculate the average, the first factor,
$\psi_{n,m}$ or $\psi_{n,m^{\prime}}$, is expressed from the main
relation eq. (\ref{recursion D}), while the second factor,
$\psi_{n-1,m}$ or $\psi_{n-1,m^{\prime}}$, is left unchanged.
Analogously, the equation for the moments $a_{n,s}$ and $c_{n,s}$
are obtained. We have ($n=2,3,\ldots$)
\begin{eqnarray}\label{Ma1}
a_{n,s}= -c_{n,s}+\mathcal{L}b_{n,s}, s\neq 0 ,\\\label{Mb1}
b_{n,s}= -b_{n-1,s}+\mathcal{L}a_{n-1,s} ,\\\label{Mc1} c_{n,s}
=-a_{n-2,s}+\mathcal{L}b_{n-1,s}.
\end{eqnarray}
The operator $\mathcal{L} $ is introduced in eq.(\ref{L}) which
acts on the index $s$ according to the equation
\begin{equation}\label{L1}
\mathcal{L}f_{n,s}^{\nu} = E f_{n,s}^{\nu} - \sum_{\mu=\pm 1}f
_{n,s+\mu}^{\nu}.
\end{equation}
In the derivation of eqs. (\ref{Ma1})-(\ref{Mc1}) the mean of the
product of uncorrelated quantities was replaced by the product of
the means.

The basic equation for the variable $x_n$ is obtained by squaring
both sides of eq. (\ref{recursion D}) and averaging over the
ensemble. One gets
\begin{eqnarray}\label{basic}
x_n=\sigma^2 x_{n-1} +
x_{n-2}+\chi_{n-1} ,\\
\chi_n=\mathcal{L}^2a_{n,0}-2\mathcal{L}b_{n,0} .
\end{eqnarray}
The expression for $\chi_n $ includes the seconds moments of the
type $a_{n,s} $,$b_{n,s} $ introduced earlier (moments $c_{n,s} $
are obviously absent).

\subsection{Z-transform and Fourier transform}

In the following derivations we utilize two types of algebraic
transforms: the Z-transform \cite{Weiss}
\begin{equation}\label{Z}
X=\sum_{n=0}^\infty \frac{x_n}{z^n},\quad
F_s^{\nu}=\sum_{n=0}^\infty \frac{f_{n,s}^{\nu}}{z^n},
\end{equation}
and the discrete Fourier transform
\begin{equation}\label{FT}
F_{s}^{\nu}=\frac 1{ 2\pi }\int_{-\pi}^{\pi} F^{\nu}(k)e^{iks}d k.
\end{equation}
Z-transform of the eqs. (\ref{Ma1})-(\ref{Mc1}) gives
\begin{eqnarray}\label{Ma2}
A_{s}= -C_{s}+\mathcal{L}B_{s}+z^{-1}\Gamma_s , s\neq 0
,\\\label{Mb2} B_{s}= -z^{-1}B_{s}+z^{-1}\mathcal{L}A_{s}
,\\\label{Mc2} C_{s} =-z^{-2}A_{s}+z^{-1}\mathcal{L}B_{s}.
\end{eqnarray}
For boundary condition, eq.(\ref{boundary}), we have
\begin{equation}\label{A0}
  A_0=X .
\end{equation}
After simplification of the equation for the moment $C_s$ we have
eq.(\ref{Mb2}) and new equation for moment $A_s$:
\begin{eqnarray}\label{Ma3}
A_{s}(1-z^{-2})= (1-z^{-1})\mathcal{L}B_{s}+z^{-1}\Gamma_s , s\neq
0 .
\end{eqnarray}
It turns out to be convenient to lift the constraint $\quad s
\neq 0$ in the form:
\begin{eqnarray}\label{Ma4}
A_{s}(1-z^{-2})= (1-z^{-1})\mathcal{L}B_{s}+z^{-1}\Gamma_s
+R\delta_{s,0} .
\end{eqnarray}
This equation requires expressing the parameter $R$ in a
self-consistent way via the boundary condition, eq.(\ref{A0}).

After an additional Fourier transform one gets
\begin{eqnarray}\label{Ma5}
A(k)(1-z^{-2})= (1-z^{-1})\mathcal{E}(k)B(k)+R+z^{-1}\Gamma (k)
,\\\label{Mb5} B(k)(1+z^{-1})= z^{-1}\mathcal{E}(k)A(k) ,
\end{eqnarray}
Here
\begin{eqnarray}
\mathcal{E}(k)=E-2\cos(k)
\end{eqnarray}
is the Fourier transform of the operator $\mathcal{L}$.

Z-transform of the basic equation (\ref{basic}) gives
\begin{eqnarray}\label{basic Z}
X=z^{-1}\sigma^2 X +
z^{-2}X+z^{-1}\chi ,\\
\chi=\mathcal{L}^2A_{0}-2\mathcal{L}B_{0}+\Gamma_0 .
\end{eqnarray}
The function $\chi$ can be represented with the help of the Fourier transform
\begin{eqnarray}\label{Chi1}
\chi=\frac{1}{2\pi}\int_{-\pi}^{\pi} \left
\{\mathcal{E}(k)^2A(k)-2\mathcal{E}(k)B(k)+\Gamma (k) \right \} dk
.
\end{eqnarray}
After simplification of eq.(\ref{Chi1}) (we use here
eqs.(\ref{Ma5}),(\ref{Mb5})) one gets
\begin{eqnarray}\label{Chi2}
\chi=z(1-z^{-2}) \frac{1}{2\pi}\int_{-\pi}^{\pi} A(k) dk -zR,
\end{eqnarray}
or
\begin{eqnarray}\label{Chi3}
\chi=z(1-z^{-2}) X -zR,
\end{eqnarray}
because for the boundary condition, eq.(\ref{A0}), we have
\begin{eqnarray}\label{iden}
A_0= \frac{1}{2\pi}\int_{-\pi}^{\pi} A(k) dk = X .
\end{eqnarray}
>From eq.(\ref{basic Z}) together with eq.(\ref{Chi3}) we obtain
\begin{eqnarray}\label{recurs3}
R=z^{-1}\sigma^2 X .
\end{eqnarray}

It is apparent, that the moment $A(k)$ as a solution of the eqs.
(\ref{Ma5}),(\ref{Mb5}) depends on the parameter $R$. The
parameter $R$ in turn is determined by $X$ according to eq.
(\ref{recurs3}). One gets
\begin{eqnarray}\label{A}
A(k) \left \{ w^2-\mathcal{E}^2(k) \right \}=\frac{(z+1)}{(z-1)}
\left \{\Gamma (k)+ \sigma^2 X \right \} ,\\
w^2=\frac{(z+1)^2}{z} .
\end{eqnarray}
The equation for the signal $X$ is obtained in a self-consistent way from
eq.(\ref{iden}). Then we finally obtain eq.(\ref{XH}), where
\begin{eqnarray}\label{X0}
X^{(0)}(z)= \frac{(z+1)}{(z-1)} \frac{1}{2\pi}\int_{-\pi }^\pi
\frac{\Gamma (k) dk}{w ^2-{\mathcal{E}}^2 (k)},
\\ H^{-1}(z)=1-\frac{\sigma ^2 (z+1)}{2\pi (z-1)
}\int_{-\pi }^\pi \frac{dk}{w ^2-{\mathcal{E}}^2 (k)}.\label{H}
\end{eqnarray}
Note that the boundary conditions (field $\alpha_m$ or correlation
function $\Gamma_s$) influence only the function $X^{(0)}(z)$
which is independent of the parameter $\sigma$ and describes an
ideal system, $H(z)=1$ for $\sigma=0$.

This point needs some comments. We have already stated
that the initial conditions $\alpha_m$ only
determine the signals, the filter is a fundamental function of the
disorder $\sigma$ only. This fact has a very simple and important
logical consequence. A filter $H(z)$ is defined  via the relation
between input and output signals, eq.(\ref{XH}). It is therefore
completely sufficient to determine only once this relation
e.g. for a particular boundary condition,
$\alpha_m$, where the calculation of input and output signals is trivial.
The simplest case is $\alpha_m=\alpha=const $. For this condition
our system has (after averaging  over random
potentials) the translational invariance along the $m-$axis
and we need not do a further averaging over the
ensemble of translations in $m$-space. For $\alpha_m=\alpha$
we get back to eq.(\ref{H}). The corresponding input signal
$X^{(0)}(z)$ follows from eq.(\ref{X0}), if one makes use of the simple
relation $\Gamma_s=\alpha^2$ .

It is appropriate to return to the topic of averaging over the ensemble of
translations in $m$-space and the full averaging, section
\ref{signal}. We clearly see now that this procedure is not at all obligatory.
One could avoid it altogether. We
have, however, used this procedure for pedagogical reasons
to demonstrate clearly that the mentioned property of the filter
(the filter is a function of the disorder $\sigma$ only)
really exists and is independent of the initial conditions.

\section{Results}

\subsection{Case $E=0$}

For the sake of illustration we restrict ourselves first to the
case of the band center $E=0$. Evaluating integrals by standard
methods (contour integrals) one gets:
\begin{equation}\label{Int}
\frac{1}{2\pi }\int_{-\pi }^\pi \frac{dk}{w ^2-{\mathcal{E}}^2
(k)} =\frac 1{w \sqrt{w ^2 -4}},
\end{equation}
where generally the complex parameter $w = u +i v$ is defined in
the upper half-plane, $v = Im (w) \geq 0$. Changing the complex
variable $z$ to the parameter $w$ corresponds to the conformal
mapping of the inner part ($|z| \leq 1$, transform $w =
-(z^{1/2}+z^{-1/2})$) or the outer part ($|z| \geq 1$, transform
$w = (z^{1/2}+z^{-1/2})$) of the circle onto the upper half-plane,
the circle itself maps onto the interval $ [ -2,2 ]$. Note also
that if $H(z)$ has complex conjugate poles, then on the upper $w$
half-plane they differ only by the sign of $u=Re (w)$. To avoid
complicated notations, we seek for poles in the sector $u\geq 0, v
\geq 0$ and double their number if we find any. The inverse
function
\begin{equation}\label{zw}
z=-1+\frac{w ^2}2\pm \frac w 2\sqrt{w ^2-4}
\end{equation}
is double-valued, it has two single-valued branches that map the
selected $w$ sector onto  either the inner part of the half-circle
($|z| \leq 1$, $(-)$ sign in the formula) or the half-plane with
the half-circle excluded ($|z| \geq 1$, $(+)$ sign in the
formula). It is also easy to derive that
\begin{equation}\label{gpm}
\frac{(z+1)}{(z-1)}=\pm \frac{w}{\sqrt{w ^2-4}}.
\end{equation}
Substituting (\ref{Int}) and (\ref{gpm}) into  (\ref{H}) gives the
system function
\begin{equation}\label{qpmw}
H^{-1}_{\pm}(w)=1 \mp \frac{\sigma ^2 }{w ^2-4} ,
\end{equation}
or
\begin{equation}\label{qpm}
H^{-1}_{\pm}(z)=1 \mp \frac{\sigma ^2 z}{(z-1)^2} .
\end{equation}
We see that the filter $H(z)$ is a non-analytic function of the
complex variable $z$. The unit circle $|z|=1$ divides the complex
plane into two analytic domains: the interior and exterior of the
unit circle. The inverse Z-transform is quite generally defined
via countour integrals in the complex plane, eq.(\ref{count}), and
this definition is only possible in an analytic domain. In this
way in the formal analysis of the problem multiple solutions
result.

Let us consider first the solution $H_+(z)$ which is formally
defined in the region $|z| \geq 1$. The function $H_+(z)$ has two
poles $\lambda_1=\lambda$ and $\lambda_2=\lambda^{-1}$, where
$\lambda=\exp(2\gamma)$, $2\sinh(\gamma)=\sigma$ and
\begin{equation}\label{gam}
\gamma=\sinh^{-1}(\frac{\sigma}{2}) .
\end{equation}
The first pole lies inside the region of definition and the second
one is located outside of it (virtual pole). However, for the
inverse Z-transform this fact is irrelevant. Note also that in the
case of general $E$ the pole $\lambda_1$ which lies on the real
axis and which can be found from the parametric representation,
eq.(\ref{zw}), in the $w$ sector defined earlier, has its virtual
counterpart $\lambda_2=\lambda^{-1}_1$. For $\sigma>0$ the ROC for
a causal filter is given by the inequality $|z|>\lambda>1$. Hence,
the unit circle does not belong to the ROC, and therefore the
filter $H_{+}(z)$ is unstable. The inverse Z-transform
gives
\begin{equation}\label{h+}
h^{+}_n=\delta _{n,0}+2\tanh (\gamma )\sinh (2\gamma n) ,
\end{equation}
which is an exponentially growing function. This result can be generalized for
all $E$ values, however the expression for the function $H_{+}(z)$
will be more complex.

Therefore, the solution given by the system function $H_+(z)$
always gives unbounded sequences $x_n$, in full analogy with the
solution of the one-dimensional problem \cite{Molinari}. The
natural interpretation of this result is that states are
localized.  The parameter $\gamma$ is nothing else but the
generalized Lyapunov exponent \cite{Molinari}, defined for the
second moment of the random amplitudes. Therefore, the
localization length is $\xi=\gamma^{-1}$.

The case of the filter $H_{-}(z)$ is a little more complicated.
The filter is formally defined in the region $|z| \leq 1$. For
$E=0$ and $\sigma >2$ the poles are found at
$\lambda_1=-\lambda^{-1}$ and $\lambda_2=-\lambda$ (virtual pole)
with $\lambda=\exp(2\gamma^{\prime })$, $2\cosh(\gamma^{\prime
})=\sigma$. A ROC which is consistent with the causality restriction
corresponds to the inequality
$|z|>\lambda >1 $ and this lies outside the region of definition
of the solution $|z| \leq 1$. Therefore, any physically feasible
solution is absent.

However, for $\sigma <\sigma_0=2$ the poles lie on the unit circle
$\lambda_{1,2}=\exp(\pm 2i\varphi)$, $2 \sin(\varphi)=\sigma$. The
critical value $\sigma_0=2$ corresponds to the equation
\begin{equation}\label{Hw}
H^{-1}_{-}(w=0)=0 .
\end{equation}
Let us consider this problem as a limiting case of a modified
problem in which the poles are shifted into the unit circle,
$\lambda^{\prime }_{1,2}=\lambda_{1,2}\exp(-\eta)$ and $\eta
\rightarrow +0$ . The casual filter has ROC $1\geq
|z|>\exp{(-\eta)})$ (taking into account the region of
definition). The ROC includes the unit circle, the filter is thus
stable. In the limit $\eta \rightarrow +0$, however, the poles
move onto the unit circle. In the literature on electrical
engineering \cite{Weiss} the filter that has a pole on the unit
circle, but none outside it is called
marginally stable. Marginally stable means there is a
bounded input signal $x_n^{(0)}$ that will cause the output $x_n$
to oscillate forever. The oscillations will not grow or decrease
in amplitude. The inverse Z-transform gives
\begin{equation}\label{h-}
h^{-}_n=\delta _{n,0}+2\tan  (\varphi )\sin (2\varphi n) ,
\end{equation}
i.e. a bounded oscillating function.

\subsection{Case $E\neq 0$}
Evaluating the required integrals one gets
\begin{eqnarray}\label{Int2}
H^{-1}_{\pm}(w)=1 \mp \frac{\sigma ^2 }{2 \sqrt{w ^2-4}}\times\\
\nonumber \left\{ \frac{1}{\sqrt{(w+E)^2
-4}}+\frac{1}{\sqrt{(w-E)^2 -4}} \right\} .
\end{eqnarray}
A similar analysis as above shows that the
marginally stable filter exists for $\sigma < \sigma_{0}(E)$,
where
\begin{equation}\label{Esigma}
\sigma_{0}(E) = 2 (1-E^2/4)^{1/4} ,
\end{equation}
and the value of the energy is limited by $\left| E\right| \leq
2$. For either $\left| E\right| > 2$ ($\sigma \neq 0$) or $\sigma
> \sigma_{0}(E)$ ($\left| E\right| < 2$) a physical  solution of
this filter is absent. The critical value of $\sigma_{0}(E)$ follows again
from eq.(\ref{Hw}).

We conclude that the system function $H_{-}(z)$  exists in a
well-defined region of energies $E$ and disorder $\sigma$. It
always gives bounded sequences $x_n$ which can be naturally
interpreted as delocalized states. Marginal stability corresponds
to the existence of quite irregular wave functions.

The first solution $H_{+}(z)$ is defined outside the unit circle
and always exists. The filter $H_{+}(z)$ describes localized
states and it is possible to connect its properties with the
notion of the localization length. In the energy range
$0<|E|<4$ the Lyapunov exponent is a non-analytical function at zero
disorder: $\lim_{\sigma \rightarrow +0}\gamma(\sigma) \neq
\gamma(0)$, because without disorder all states here are extended ones,
$\gamma(0)=0$. The second solution $H_{-}(z)$ is
defined inside the unit circle and does not always represent a
solution which can be physically interpreted (this is the
mathematical consequence that the filter be causal). The filter
$H_{-}(z)$ describes delocalized states.

We also note that if both solutions $h^{+}_n$ and $h^{-}_n$ for
the system function exist simultaneously, they both give solutions
$x_n$ of the initial problem that satisfy all boundary conditions.
In this sense, a general solution of the problem is $h_n=\omega
h^{+}_n+(1-\omega)h^{-}_n$, where the parameter $\omega$ is left
undefined by the averaging procedure (this is natural for the
problem with particular solutions of different asymptotic
behavior). Such a solution $h_n$ may be interpreted as
representing two phases, considering that $h^{+}_n$ determines the
properties of the insulating state, but $h^{-}_n$ the metallic
state. Therefore the metal-insulator transition should be looked
at from the basis of first-order phase transition theory. This
opinion differs from the traditional point of view, which
considers this transition as continuous (second-order). For first
order phase transitions the coexistence of phases is a general
property. There exists for the present case an experimental result which is
at least consistent with the present non-trivial result. Ilani et al.\cite{Ilani1,Ilani2}
have studied the spatial structure at the metal-insulator
transition in two dimensions. They found \cite{Ilani1}: 'The
measurement show that as we approach the transition from the
metallic side, a new phase emerges that consists of weakly coupled
fragments of the two-dimensional system. These fragments consist
of localized charge that coexists with the surrounding metallic
phase. As the density is lowered into the insulating phase, the
number of fragments increases on account of the disapearing
metallic phase.'

\subsection{Perturbation theory}

The results of the present investigation - if it proves to be correct -
have certain consequences
for the validity of many theoretical tools used for the investigation of Anderson
localization in the past. Our comments do not exclude the possibility
that better theroretical tools in the traditional approaches can be found,
which may better master the situation. Our comments in the subsequent
paragraphs are intended to highlight the differences between the present
and previous approaches and the basic problems encountered by previous
approaches - all this under the premise that the present analytical theory
is proven to be correct. Our first comment refers to the Anderson
localization as a critical phenomenon. One has always believed
that all states in a 1-D and 2-D systems are localized for
infinitesimal disorder, whereas in 3-D a metal-insulator
transition occurs. From our studies it, however, emerges that a
metal-insulator transition occurs in 2-D system.

Because localization as a metal-insulator transition is a typical
critical phenomenon, the parameters $\gamma$ and  $\xi$ are quite
generally not analytical functions neither in the energy nor in
the disorder parameter. It is well-known that perturbation theory
is not a suitable method for describing critical phenomena.

E.g. for $E=0$  and $\sigma \rightarrow 0$ one has from
eq.(\ref{gam}) $\gamma \propto \sigma$. From this follows that the
function $\gamma$ cannot be represented as a series in powers of
$\sigma^2$. I.e. perturbation theory is not applicable to the
Anderson problem in 2-D (as it is neither for other critical
phenomena \cite{Baxter}): the corresponding series expansions
tend to diverge. This should be quite generally valid; that is why
all estimations which result from first order perturbation theory
(e.g. for the mean free path) are physically extremely doubtful,
because they are the first term of a divergent series.

In the case of 1-D systems it is
known\cite{Kramer,Kappus,Molinari} that in the limit of small
disorder all Lyapunov exponents $\gamma \propto \sigma^2$. Here
perturbation theory is completely acceptable, because 1-D systems
do not show a phase transition \cite{Stanley}. These results of
perturbation theory tend, however, to become inapplicable for higher spatial
dimensions. In a similar vein it can be stated that results for the 1-D Ising-model have
no relevance for the 2-D model (Onsager solution)
\cite{Stanley,Baxter}.

\subsection{Order of phase transition and self-averaging}

Physics of disorder associates experimental quantities with quantities obtained by
averaging over random potentials (statistical ensemble of
macroscopically different systems). This approach is well-known \cite{Kramer}.
One expects that, although the results
of measurement of this physical quantity are dependent on the
realization of disorder, the statistical fluctuations of the
result are small. One assumes that physical quantities are only those
which do not fluctuate within the statistical ensemble in
the thermodynamic limit (length of system $L \rightarrow \infty$).
One defines these as self-averaging quantities. It is also known
that in connection with localization this
self-averaging property is not trivially fulfilled, e.g. the
transport properties of the disordered systems are in general not
self-averaging\cite{Kramer}. Here the fluctuations are much larger
than expected or they even diverge in the thermodynamic limit.

Even though the existence of non-self-averaging quantities is known,
the theory of the physics of disorder starts from the assumption that
certain quantities will still be self-averaging and physical. However, one
has to the authors' knowledge never analyzed the conditions for
the validity of this statement from
the point of view of phase transition theory. In a system without
phase transitions one can prove the existence of certain self-averaging
quantities. Their existence is also possible if there is a second order
phase transition. Here in the thermodynamic limit there exists only one
or the other phase, a coexistence of phases is impossible by principle.
In each phase there do exist certain average values which have physical relevance.
Strong fluctuations are observed only as usual in the vicinity of a critical point.
In a system with a first order phase transition one finds
certain parameter values for which two phases do coexist; i.e. a
macroscopic system is heterogeneous and consists of
macroscopically homogeneous domains of the two phases. In this
case a formal averaging over the statistical ensemble takes into
consideration also an averaging over the phases, and the resulting
averages have no physical meaning. Physically meaningful are only
the properties of the pure homogeneous phases. The trivial example
in this respect is the coexistence of water and ice. An average
density of this system has no sense and depends on the relative
fraction of the phases, whereas the density of pure water and pure
ice are meaningful quantities.

Formally systems with a first order phase transition do not
possess self-averaging quantities. The well-known idea
\cite{Kramer}, to analyze such non-self-averaging quantities via
probability distribution functions or all of its moments, cannot
be realized in practice. In the theory of phase transitions
\cite{Stanley} one has a general idea to treat such
multi-component systems, but these exist only as approximations.
One writes down the equations for certain averages. If these
equations possess a multiplicity of solutions one interprets the
corresponding solutions as phases. Let us now consider the present
solution of the Anderson problem from this point of view.

It is well-known that the exact equations of statistical physics
are always linear but form an infinite chain (e.g. see the
equations which form the basis of the BBGKY theory
(Bogoliubov, Born, Green, Kirkwood, Yvonne) \cite{Balescu}.
Via decoupling approximations which use multiplicative forms of
distribution functions (see e.g. the Kirkwood approximation
\cite{Balescu}) one derives from the insoluble infinite chain a
finite set of equations which, however, are now nonlinear. This
latter property leads to the multiplicity of solutions and  the
possibility to describe the phase transition. It remains, however,
unclear in which way the original linear equations describe
mathematically the phase transition. The Anderson localization
problem is the extremely rare case where an analytic solution was
found for the chain of equations. As a result one can clearly see
how mathematics produces a multiplicity of solutions in the form of
non-analytic filter functions $H(z)$.
The Z-transform lifts the phase degeneracy allowing one to
study the properties of each phase independently. Particular
solutions $h^{+}_n$ and $h^{-}_n$ can be naturally interpreted not
as the result of full averaging over all realizations of random
potentials, but as averaged over one of the two classes of
filtered realizations.
If a particular realization of random potentials for a
given energy $E$ leads to unbounded sequences for all boundary
conditions, we classify it as belonging to the first class.
Otherwise the realization of random potentials belongs to the
second class. Sorting of the realizations into two classes
corresponds mathematically to an infinitely weak constraint on the
on-site potentials, which are still characterized in the equations
by a single parameter $\sigma$. Therefore the equations for the
second moments given above will describe both one-phase solutions
with full averaging (if the metallic phase is impossible) and each
phase of the two-phase solution separately with partial
averaging.

\subsection{Numerical scaling}

For known technical reasons one treats D-dimensional systems as
quasi-one-dimensional ones \cite{Kramer};
this is e.g. done for the recursive calculation of the Lyapunov
exponents. One starts with a system of the size $\infty \times L^{D-1}$:
It is infinite in the $n-$direction and the length $L$ in the other
directions is finite. Numerical scaling studies \cite{Kramer}
assume the scaling variable to be self-averaging. A further crucial
assumption is that there is one-parameter scale function.

One calculates the localization length $\xi_L$ as
the average of the logarithm of the quantum mechanical
transmission probability \cite{Kramer}, and assumes that this
quantity is self-averaging. For any finite length $L$ the
localization length $\xi_L$ is finite too. This fact is rather
trivial. Every quasi-one-dimensional system is
qualitatively similar to a true one-dimensional one. Because in the
1-D Anderson model all states are localized, the same will hold for
quasi-one-dimension systems. It is known that a rigorous
theoretical definition of the phases requires the investigation
of the thermodynamic limit $L \rightarrow \infty$. One assumes
that in order to be extrapolate to infinite system size
it is necessary to investigate the scaling behaviour of
$\xi_L$\cite{Kramer}. It is possible to establish a scaling
function
\begin{equation}\label{scaling}
\frac{\xi_L}{L}=f(\frac{\xi_{\infty}}{L})
\end{equation}
that does not depend on the disorder. The scaling parameter
$\xi_{\infty}$ is a function of disorder. This method yields
complete localization in 2-D dimension, and an Anderson
transition in 3-D case.
\begin{figure}[htbp]
  \begin{center}
    \includegraphics[width=8cm]{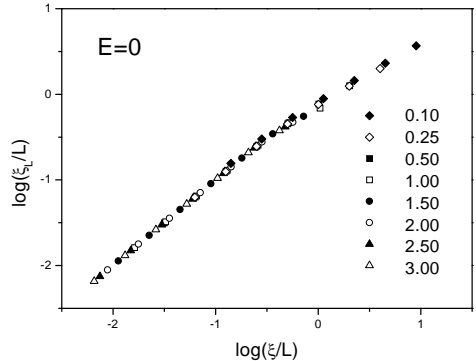}
    \caption{
     Scaling function (for the second moment) for the 2-D Anderson model.
     $\xi=\xi (\sigma)=\xi_{\infty}$ is the scaling parameter necessary to
     scale numerical data onto the same curve. Values of the
     disorder $\sigma$ are indicated.
      }
    \label{fig1}
  \end{center}
\end{figure}

One notices that the idea of numerical scaling has
a weak point; one starts from the assumption that self-averaging
quantities exist. The analytic solution presented in the
present paper, section \ref{2D}, offers in principle the possibility
to clarify the problems connected with the numerical scaling procedure.
The analytic equations treat from the very start a system in the
thermodynamic limit ($L=\infty$), because the integer $m \in
(-\infty , +\infty )$ in eq.(\ref{schroedinger}). A transition to a
finite system ($L<\infty$) in this context is rather trivial and
corresponds simply to the transition from the Fourier integral,
eq.(\ref{FT}) to the Fourier series.
Thus we arrive from the filter-function in the form of eq.(\ref{H})
at another equation:
\begin{eqnarray}\label{HL}
H^{-1}(z)=1-\frac{\sigma ^2}{L}
\frac{(z+1)}{(z-1)}\sum_{j=0}^{L-1}
\frac{1}{w^2-{\mathcal{E}}^2 (k_j)} ,\\
k_j=\frac{2\pi j}{L} .
\end{eqnarray}
The difference is large. The filter (\ref{H}) has a multiplicity
of solutions: the unit circle $|z|=1$ divides the complex plane
into two analytic domains. For a finite system one has only
one solution, the unit circle $|z|=1$ does not play a role anymore.
The filter (\ref{HL}) corresponds to only one phase, where all states
are localized. Mathematically this means that between the possible poles
$z=\lambda_i$ of the function (\ref{HL}) there always exist the pole
$\lambda_{max}>1$, and the filter is unstable. If one defines the
localization length as $\xi_L=1/\gamma_L$, where according to the definition
by using the second moments $\lambda_{max}=\exp(2\gamma_L)$, then
this length is always finite.

The physical reason for this result is rather clear:
in the $n$-direction the system remains infinite since the
$Z$-transform always ``feels'' the true asymptotic behavior.
Fixing a finite size $L$ in the $m$-direction effectively renders
the system  a one-dimensional one at large scales, hence the corresponding
behaviour (full localization).
From this point of view, $\xi_L$ is essentially the crossover length form
2-D to 1-D behavior.

Let us do now a typical scaling study for the localization length
$\xi_L$ at the band centre, $E=0$. We have found that the scaling
parameter $\xi_{\infty}$ is identical with the value $\xi$,
where $\xi=1/\gamma$ and $\gamma$ is defined in
eq.(\ref{gam}). Disregarding for the moment the fact that one
commonly uses a different definition of the localization
length $\xi_L$ via the log-definition, we arrive qualitatively at
the same results as numerical scaling\cite{Kramer}. There exists a scaling
function, Fig.1,  which is typical for 2-D systems and whose behaviour
one commonly interprets as complete localization in
2-D dimensions. The corresponding scaling parameter
is, however, identical with the localization length $\xi$ in the
insulating phase. I.e. numerical scaling is not capable to analyze a
system consisting of two phases. This approach to the problem with
particular solutions of different asymptotic behaviour,  $\gamma
\equiv 0$ or $\gamma \neq 0$, always sees only one diverging solution
($\gamma \neq 0$).

\section{Conclusion}

The basic idea of the present work is to apply signal theory to the
Anderson localization problem. We interpret certain moments of the
wave function as signals. There is then in signal theory a qualitative
difference between localized and extended states: The first ones
correspond to unbounded signals and the latter ones to bounded signals.
In the case of a metal-insulator transition extended states
(bounded signals) transform into localized states (unbounded signals).
Signal theory shows that it is possible in this case to find a
function (the system function or filter), which is responsible for
this transformation. The existence of this transformation in
a certain region of disorder and energy simply means that
the filter looses its stability in this region. The meaning of an
unstable filter is defined by a specific pole diagram in the complex
plane. These poles also define a quantitative measure of
localization. Thus it is possible here to determine the socalled
generalized Lyapunov exponents as a function of disorder and energy.

For pedagogical reasons we have analyzed in this paper first the 1-D
case. Here no new results are obtained. All states are localized for
arbitrary disorder. The aim of this section consists in showing to
the uninitiated reader a possible alternative to the traditional
mathematical tools of dealing with Anderson localization and to interpret
its content. The power of the approach comes only to its full bearing
when it proves possible to find analytically the filter function
also for the 2-D case. Although the filter is quite in general defined
by an integral, it is possible to give a representation in terms of radicals.
As a consequence we have been able to find an exact analytical solution for the
generalized Lyapunov exponents for the well-known and notorious Anderson problem in 2-D.
In this way the phase diagram is obtained.

The approach suggested in the present paper does not in any way
represent the complete solution to the problem of Anderson
localization in 2-D. The very important topic of the transport properties
of disordered systems lies outside the presumed region of applicability
of the method. The results of the studies give us a few answers
to existing questions but open up even more questions. We have only considered the
phase diagram of the system. The theory permits us to ascertain to which
class of problems Anderson localization belongs.

In the 1-D case all states are localized for infinitesimally small
disorder in complete agreement with the theoretical treatment of
Molinari \cite{Molinari}. In the 2-D case we have shown that in
principle there is the possibility that the phase of delocalized
states exists for a non-interacting electron system. This phase
has its proper existence region and it belongs to the marginal
type of stability. All states with energies $|E| > 2$ are
localized at arbitrarily weak disorder. For energies and disorder,
where extended states may exist we find a coexistence of these
localized and extended states. Thus the Anderson transition should
be regarded as a first order phase transition. The argument of
Mott that extended and localized states cannot be found at the
same energy is probably not applicable in our case as the solutions for the
two relevant filter functions do not mix. This has to be worked out
in detail in the future. Also the role of different disorder realizations has
to be elaborated, which could be a source of the observed phenomena.

Our findings can in principle explain the experimental results of
Kravchenko et al. \cite{Kravchenko} on films of a-Si. They are also in
agreement with the experimental findings of Ilani et al \cite{Ilani1,Ilani2} which
showed the coexistence of phases. Although one has to be cautious in
concluding from our results directly onto specific experimental
situations our theory is the first one to be in accord with experiment
for 2-D systems.

We hope that the presented results contribute to the solution of the existing
contradictions between theory and experiment, and also between
different experimental results.

The qualitative diagnosis which we propose here is in several aspects different
from the predictions of prevailing theories, scaling theory, mean field
and perturbation theory. It has to be stated here that much work still remains to
be done both via the present approach as via other established approaches.

\begin{enumerate}
\item The Anderson problem belongs to the class of critical
phenomena. It is known from the scientific literature that in
such a case one encounters in general a poor convergence behaviour
of approximations \cite{Stanley,Baxter}, which in their basis
contain a mean field approach. To cite an example: the exact
solution of Onsager for the 2-D Ising model is in complete
disagreement with the Landau theory of phase transitions (mean
field theory) \cite{Stanley}. All critical exponents differ in the two
approaches. Yet there is at least a qualitative agreement between
the two theories: the Landau theory describes second order phase
transitions and the exact Onsager theory confirms this aspect.

\item  For the
Anderson problem the outlook on the theories might also be
far reaching, if our approach should prove to be correct. All
presently accepted theories derive here from the scaling idea
which is in turn based on the assumption that the Anderson
transition belongs to the class of second order phase transitions.
This basic idea is up to now a hypothesis which cannot be proved
in the absence of the exact solution. One has assumed that
averaged or self-averaging quantities are always physical
quantities. And every theory had the direct aim to calculate such
quantities. Our exact solution for the phase diagram, however,
indicates that in this system the coexistence of phases is
possible, i.e. that the Anderson transition should be regarded as
a phase transition of first order.

\item
For a phase transition of first order with a coexistence of phases there surely
also exist self-averaging quantities, whose fluctuations are small. The quantities,
however, tend to be unphysical because the average over an ensemble
includes an average over the phases. Physically meaningful in this
case are only the properties of the pure phases. Consequently one requires a
new idea to calculate the transport properties of the disordered systems. We
can at the present time only surmise that this future theory of transport properties will
calculate the transport properties with the help of equations which have the
property of a multiplicity of solutions. This theory should also formulate
in a new way the average over the ensemble of random potentials in order to
arrive at the properties of the pure phases.

\end{enumerate}

\ack{V.N.K. gratefully acknowledges the support of the Deutsche
Forschungsgemeinschaft. This work was partly supported by the EC
Excellence Centre of Advanced Material Research and Technology
(contract N 1CA1-CT-2080-7007) and the Fonds der Chemischen
Industrie. We thank Dr. L. Schweitzer and colleagues very much for
illuminating discussions.}

\end{document}